\pdfoutput=1

\documentclass[11pt]{article}

\usepackage[preprint]{acl}

\usepackage{times}
\usepackage{latexsym}
\usepackage{amssymb}

\usepackage[T1]{fontenc}

\usepackage[utf8]{inputenc}

\usepackage{microtype}

\usepackage{inconsolata}

\usepackage{graphicx}
\usepackage{amsmath}
\usepackage{booktabs}
\usepackage{color}
\usepackage{hyperref}

\title{{Fast-MIA}: Efficient and Scalable Membership Inference for LLMs}

\author{Hiromu Takahashi\\
  Independent Researcher\thanks{This work was conducted as part of a business outsourcing agreement with Nikkei Inc.} \\
  Tokyo, Japan \\
  \texttt{hiromu.takahashi56@gmail.com} \\\And
  Shotaro Ishihara\thanks{Corresponding author}\\
  Nikkei Inc.\\
  Tokyo, Japan \\
  \texttt{shotaro.ishihara@nex.nikkei.com} \\}

\begin{document}
\maketitle
\begin{abstract}
We propose Fast-MIA (\url{https://github.com/Nikkei/fast-mia}), a Python library for efficiently evaluating membership inference attacks (MIA) against large language models (LLMs).
MIA has emerged as a crucial technique for auditing privacy risks and copyright infringement in LLMs. 
However, computational demands have grown substantially: recent methods rely on repeated inference, while practical auditing requires large-scale evaluation.
Progress is further hindered by existing implementations that execute methods independently, redundantly computing shared intermediate results such as log-probabilities.
To address these challenges, Fast-MIA combines two strategies: (1) high-throughput batch inference via vLLM, achieving approximately 5$\times$ speedup, and (2) a cross-method caching architecture that computes intermediate results once and shares them across methods.
The library includes representative MIA methods under a unified framework, integrates with established benchmarks, and supports flexible YAML configuration.
We release Fast-MIA under the Apache License 2.0 to support scalable and reproducible MIA research.
\end{abstract}

\section{Introduction}

As LLMs become increasingly deployed in practical applications, concerns have emerged regarding their tendency to \emph{memorize} training data~\cite{ishihara-2023-training}. 
Memorization refers to the phenomenon in which a model reproduces exact or nearly identical sequences from its training corpus. 
This behavior can lead to privacy violations, copyright infringement, and diminished generalization capabilities.
In the context of privacy, an early study by \citet{274574} demonstrated that personal information could be extracted from GPT-2~\cite{radford2019language}. 
With respect to copyright,~\citet{10.1145/3543507.3583199} investigated GPT-2 and raised ethical concerns surrounding potential plagiarism. 
Moreover, there is growing concern about \emph{data contamination}, where memorization of benchmark datasets by LLMs may compromise the reliability of model evaluations~\cite{magar-schwartz-2022-data}.

To assess such risks of LLMs, \emph{membership inference attacks (MIA)}~\cite{7958568} have been widely adopted.
These attacks on LLMs aim to determine whether a particular data instance is included in the training set.
In the context of LLMs, MIA has emerged as a key research area due to growing demands for transparency and accountability in AI systems~\cite{Wu2025-ar}.

However, the computational requirements for MIA evaluation have grown 
substantially, hindering both research and practical implementation.
Recent advances in MIA methodology increasingly rely on 
repeated inference: text perturbation methods such as ReCaLL~\cite{xie-etal-2024-recall} and Con-ReCall~\cite{wang-etal-2025-con} require computing losses across multiple prefix configurations, while black-box approaches like SaMIA~\cite{kaneko-etal-2025-sampling} demand numerous generation steps per sample.
Furthermore, there is a growing recognition that large-scale evaluation is essential for practical copyright auditing. 
\citet{puerto-etal-2025-scaling} demonstrate that MIA becomes effective only when aggregating results across multiple documents rather than evaluating individual sentences, shifting the focus from sentence-level to collection-level membership inference.
These trends, multi-pass inference methods and dataset-scale evaluation, have made computational efficiency a critical bottleneck for MIA research.

\begin{figure}[t]
\centering
\includegraphics[width=4.5cm]{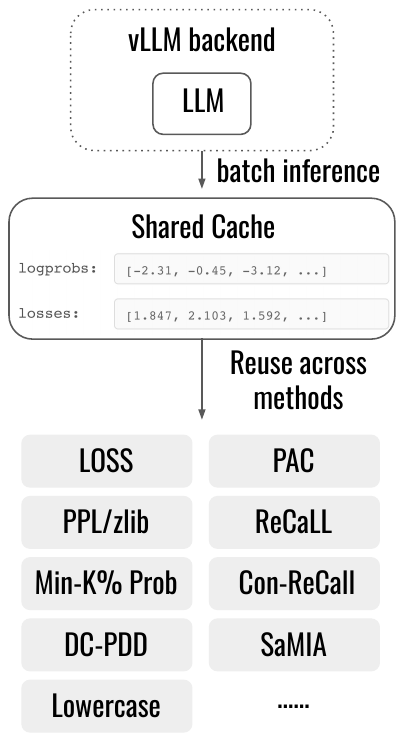}
\caption{
  Fast-MIA provides acceleration through high-throughput batch inference by vLLM and cross-method caching architecture across multiple MIA methods.
}
\label{fig:project_overview}
\end{figure}

To address these challenges, we propose Fast-MIA, an open-source Python 
library designed for efficient and reproducible evaluation of MIA against LLMs (Figure~\ref{fig:project_overview}).
Fast-MIA combines two complementary strategies for computational efficiency.
First, it provides \emph{high-throughput batch inference} by leveraging vLLM~\cite{10.1145/3600006.3613165}, achieving approximately five times faster inference compared to standard Transformers-based implementations~\cite{wolf-etal-2020-transformers} with almost no change in evaluation results. 
Second, it introduces a \emph{cross-method caching architecture} specifically designed for MIA workflows: intermediate results such as log-probabilities and losses are computed once and shared across multiple evaluation methods. 
This design is particularly beneficial for (1) fair comparison of multiple MIA methods under identical conditions, (2) hyperparameter sweeps such as varying $k$ in Min-K\% Prob, and so on.

Fast-MIA offers implementations of several representative MIA methods within a unified and extensible evaluation framework.
Users can specify models, datasets, and MIA methods using simple YAML configuration files. 
The library integrates directly with established benchmarks including 
WikiMIA~\cite{shi2024detecting} and MIMIR~\cite{duan2024do}, supports multiple data formats, and can handle languages other than English. 
Furthermore, the modular architecture enables plug-and-play extensibility for custom attack methods.
Unless otherwise noted, this paper describes and evaluates Fast-MIA version 0.3.0\footnote{\url{https://github.com/Nikkei/fast-mia/releases/tag/v0.3.0}}.

\section{Background}
\label{sec:background}

MIA provides a general framework for studying memorization, generalization, and privacy risks in machine learning~\cite{7958568}.
Formally, an attacker is given a target model \( f \), a data point \( x \), and optionally auxiliary information, and must predict whether $x$ is included in the training set.

\subsection{Mathematical Preliminaries}

Many MIA methods against LLMs rely on the token-level generation probabilities produced by autoregressive language models. Given a text sequence \( s = c_1c_2\dots c_T \) consisting of \( T \) tokens, the model assigns the following probability:
\[
p(s) = \prod_{t=1}^{T} p(c_t \mid c_1, \dots, c_{t-1})
\]
In practice, the \emph{log-likelihood} is commonly used for numerical stability:
\[
\frac{1}{T} \log p(s) = \frac{1}{T} \sum_{t=1}^{T} \log p(c_t \mid c_1, \dots, c_{t-1})
\]
Another widely used metric is the \emph{perplexity (PPL)}, which quantifies the model’s uncertainty:
\[
\text{PPL} = \exp\left( -\frac{1}{T} \sum_{t=1}^{T} \log p(c_t \mid c_1, \dots, c_{t-1}) \right)
\]
These quantities form the foundation of many MIA methods, which seek to identify systematic differences in how models score training versus non-training examples.

\subsection{Types of MIA Methods}
\label{subsec: types}

MIA methods exploit the assumption that models behave differently on training data than on unseen data.
These behavioral differences can manifest in various forms, and accordingly, different methods focus on different signals.
We classify representative approaches to MIA methods into four main categories, following \citet{chen-etal-2025-statistical}.

\paragraph{Baseline Methods.}

A simple approach utilizing PPL has been used in the initial work~\cite{274574}.

\begin{itemize}
  \item \textbf{LOSS}~\cite{8429311}: This classical baseline uses the loss. In the context
of LLMs, the loss corresponds to PPL. A lower PPL indicates a higher likelihood, which may suggest that the input was part of the training data.
  \item \textbf{{PPL/zlib}}~\cite{274574}: This method computes the ratio between the model’s PPL and the size of the zlib-compressed version of the input. Highly redundant training samples tend to yield both low PPL and small compressed sizes.
\end{itemize}

\paragraph{Token Distribution Based Methods.}

Some research has improved MIA performance by focusing on token distribution and applying heuristic techniques.
The most common method is Min-K\% Prob, and DC-PDD is an improved version that normalizes and standardizes the generation probabilities.

\begin{itemize}
  \item \textbf{Min-K\% Prob}~\cite{shi2024detecting}: Rather than considering the overall loss, this method focuses on the least confident tokens in a sequence (i.e., those with the lowest predicted probabilities). It computes the average log-likelihood over the lowest \( K\% \) of tokens, emphasizing local uncertainty.
  \item \textbf{DC-PDD}~\cite{zhang-etal-2024-pretraining}: This method calibrates token-level likelihood scores by measuring divergence from a background token frequency distribution, reducing bias toward common words.
\end{itemize}

\paragraph{Text Alternation Based Methods.}

There are methods for observing changes in log-likelihood and PPL when a text is rephrased.
The hypothesis is that when the member sample is changed, the text is expected to undergo changes that cause it to exhibit characteristics similar to those of the non-member sample.

\begin{itemize}
  \item \textbf{Lowercase}~\cite{274574}: This method compares the PPL before and after converting all input text to lowercase. The idea is that casing and similar surface features are often memorized, so altering them may affect training examples more strongly.
  \item \textbf{PAC}~\cite{ye-etal-2024-data}: This method observes changes in the log-likelihood when two random tokens are continuously swapped.
  \item \textbf{ReCaLL}~\cite{xie-etal-2024-recall}: This method appends unrelated (non-membership) text to the prompt and measures the change in log-likelihood.
  \item \textbf{Con-ReCall}~\cite{wang-etal-2025-con}: This method adds both relevant (membership) text and irrelevant (non-membership) text to the prompt. The aim is to facilitate measuring changes in the log-likelihood.
\end{itemize}

\paragraph{Black-Box Methods.}

Some research adopts settings that do not use access to generation probabilities~\cite{dong-etal-2024-generalization, kaneko-etal-2025-sampling}.
Memorization is quantified by generating the rest of the text based on the input at the beginning of the text and measuring the similarity of the strings.

\begin{itemize}
  \item \textbf{SaMIA}~\cite{kaneko-etal-2025-sampling}: This method treats multiple outputs as text samples and measures n-gram similarity.
\end{itemize}

\subsection{Challenges in Practicing MIAs on LLMs.}
\label{subsec: challenges}

Despite the increasing interest in MIA against LLMs, there remain two major barriers to advancing this line of research: (1) growing computational demands, and (2) inefficient use of shared computations across methods.
These issues underscore the need for scalable, efficient, and reproducible evaluation pipelines for MIA against LLMs.

\paragraph{Growing computational demands.}

First, the computational cost of evaluating MIA methods on LLMs is extremely high.
Unlike traditional machine learning models, LLMs require substantial GPU memory and time.
Some text alternation-based methods require multiple inferences for a single sample.
For example, Lowercase requires two inferences before and after altering.
ReCaLL and Con-ReCall also need several inferences.
The number of additional inferences in PAC and SaMIA depends on hyperparameters.
Furthermore, recent studies have highlighted the importance of inference at the dataset level, rather than relying solely on individual sentences~\cite{NEURIPS2024_e01519b4,puerto-etal-2025-scaling}, which consequently increases the required computational cost.

\begin{table}[t]
\centering
\small
\begin{tabular}{lcc}
\toprule
Paper & vLLM & Cache \\ \midrule
\citet{Murakonda2020-yr} & \\
\citet{duan2024do} &  \\
\citet{Ravaut2024-eh} & \checkmark \\
\citet{10992321} &  \\
\citet{Koike2025-hv} &  \\
\citet{Chen2026-om} &  \\
Fast-MIA & \checkmark & \checkmark \\
\bottomrule
\end{tabular}
\caption{
    List of research papers publishing implementations that combine multiple MIA techniques.
}
\label{tab:related}
\end{table}

\paragraph{Redundant computation across methods.}

Second, the implementations of MIA methods are fragmented, heterogeneous, and rarely maintained.
Although several studies have published official implementations, many public repositories are tailored to each paper and lack unified interfaces.
These implementations are often tightly coupled to specific experiments or datasets.
This makes subsequent research complicated.

Several related initiatives have released implementations bundling MIA techniques as shown in Table~\ref{tab:related}, but none have managed to eliminate redundancy during multiple executions.
Note that LLM-Sanitize~\cite{Ravaut2024-eh} has not been updated since August 2024, potentially limiting its applicability in the evolving LLM landscape.
For example, it only supports vLLM version 0.3.3 and cannot benefit from improvements introduced in subsequent versions.
Fast-MIA version 0.3.0 uses vLLM version 0.15.1, which was released in February 2026.
The recent comprehensive comparison was conducted by \citet{chen-etal-2025-statistical}, which evaluated multiple MIA methods.
Nevertheless, their implementation remains unpublished, limiting reproducibility.

\section{Fast-MIA Library}
\label{sec:proposed}

To address the computational and infrastructural challenges shown in Section~\ref{subsec: challenges}, we introduce Fast-MIA, a Python library designed for fast, reproducible, and extensible evaluation of MIAs on LLMs.
It integrates representative MIA methods and scalable inference under a single configuration-based interface, enabling reliable comparison across models, datasets, and methods.

\subsection{Key Features}

Fast-MIA is designed around two key objectives: (1) high-throughput batch inference, and (2) cross-method caching architecture.

\paragraph{1. High-throughput batch inference.}
Fast-MIA leverages the batched inference capabilities of vLLM to reduce inference time. vLLM is an open-source inference engine optimized for LLMs, featuring efficient memory paging (PagedAttention) and support for dynamic batching~\cite{10.1145/3600006.3613165}.
As a result, a speedup compared to simple Transformers implementations can be expected.

\paragraph{2. Cross-method caching architecture.}

Fast-MIA maintains an internal cache of model outputs to eliminate redundant computations across multiple methods.
It is not merely a dependence on vLLM capabilities, but rather a design-level effort to effectively reuse the cache within the library.
Fast-MIA also converts method-specific repeated inference loops into batched operations whenever possible.
This design leads to savings in both computation time and memory usage.

\subsection{System Architecture}

From a system perspective, Fast-MIA consists of a modular pipeline that connects the following components:
\begin{itemize}
  \item \textbf{Data Loader}: Accepts multiple formats, including CSV, JSON, JSONL, Parquet, and Hugging Face datasets.
  \item \textbf{Model Loader}: Loads target LLMs (e.g., Hugging Face models\footnote{\url{https://huggingface.co/models}} or LoRA adapters~\cite{hu2022lora}) through the high-throughput inference engine vLLM. Fast-MIA can load quantized models through vLLM when the corresponding quantization format is supported by the backend\footnote{\url{https://docs.vllm.ai/en/v0.15.1/features/quantization/}}.
  \item \textbf{Evaluator}: Coordinates inference, caching, and scoring for one or more MIA methods in a batched manner. It determines which intermediate model outputs are required by each method, executes only the necessary inference calls, and reuses cached outputs whenever possible.
  \item \textbf{MIA Method Registry}: Manages modular implementations of each supported MIA method.
  \item \textbf{YAML Configuration Interface}: Encodes the entire experimental setup in a human-readable, version-controllable format.
\end{itemize}

\subsection{User Interface and Configuration}

All model, data, and method parameters are specified in a YAML configuration file, allowing for consistent and reproducible evaluation across methods and experiments.
Users can use models and datasets by specifying their Hugging Face identifiers.
Fast-MIA directly supports widely used MIA benchmarks such as WikiMIA and MIMIR, while also allowing custom datasets in common formats.
This design follows common practices in the recent LLM ecosystem, where models and datasets are often specified by their Hugging Face identifiers.

For example, the following minimal configuration evaluates the LOSS method on WikiMIA using a Hugging Face model:

\begin{quote}
\footnotesize
\begin{verbatim}
model:
  model_id: "huggyllama/llama-30b"
data:
  data_path: "swj0419/WikiMIA"
  format: "huggingface"
  text_length: 32
methods:
  - type: "loss"
\end{verbatim}
\end{quote}
This unified setup eliminates method-specific scripts and facilitates batch comparisons under identical conditions.
Once the user has a configuration file, they can run MIA with the following simple command.
\begin{quote}
\footnotesize
\begin{verbatim}
uv run --with vllm python main.py \
  --config config/sample.yaml
\end{verbatim}
\end{quote}
Additional options, such as the random seed and maximum cache size, can be specified via command-line arguments.

\paragraph{Metrics.}

Fast-MIA computes the AUC as the most commonly used evaluation metric.
Furthermore, following the advice of \citet{Carlini2022-rx}, the false positive rate at 95\% TPR (FPR@95) and the true positive rate at 5\% FPR (TPR@5) are calculated.

\paragraph{Environment.}

Fast-MIA is compatible with Linux (NVIDIA A100 GPUs)\footnote{We also provide a Google Colab example for T4 GPUs.}, and provides setup instructions for environments using \texttt{uv}\footnote{\url{https://github.com/astral-sh/uv}}.
uv is a fast Python library management tool written in Rust.
All components are decoupled and version-controlled, ensuring that the framework remains adaptable as model architectures, evaluation methods, and datasets evolve.

\paragraph{Output.}

By default, Fast-MIA saves each run in a timestamped directory, including a copy of the configuration file, summary metrics, and a human-readable report.
For more detailed analysis, users can enable the detailed report option, which additionally exports per-sample scores, execution metadata, and visualization files such as ROC curves, score distributions, and metric comparison plots.
The metadata records experimental conditions such as model, data, method parameters, timing information, git information, and cache statistics, thereby supporting reproducible benchmarking and post-hoc analysis.

\subsection{Supported Methods and Extensibility}

Fast-MIA version 0.3.0 supports nine representative MIA methods, including baseline methods (LOSS and PPL/zlib), token distribution-based methods (Min-K\% Prob and DC-PDD), text alternation-based methods (Lowercase, PAC, ReCaLL, Con-ReCall), and a black-box method (SaMIA).
We have incorporated the official implementation in compliance with the license.
The implementations of PAC and SaMIA are licensed under the MIT license, and the code of DC-PDD is provided as software under the ACL Anthology (CC BY 4.0 license).
The others are licensed under the Apache-2.0 license.
Therefore, we have licensed Fast-MIA under the Apache-2.0 license.

\paragraph{Extensibility for custom methods.}

Each MIA method is implemented in a module (e.g., \texttt{loss.py}, \texttt{recall.py}) following a common interface defined in \texttt{base.py}.
This allows users to add custom attack strategies by subclassing and registering via the method factory without modifying core components.
Users can implement a custom MIA method by three steps:
(1) Creating a file for a custom MIA method (e.g. \texttt{src/methods/\{custom\_method\}.py}), (2) Implementing a new class by inheriting the \texttt{BaseMethod} class located in \texttt{src/methods/base.py} (two functions, \texttt{process\_output} and \texttt{run}, are included), and (3) Editing \texttt{src/methods/factory.py} to incorporate the custom MIA method you created.

\paragraph{Language.}

Furthermore, Fast-MIA provides a flag (\texttt{space\_delimited\_language}) for languages that are not separated by spaces, such as Japanese and Chinese.
Some research reports the trends in MIA methods that differ from English~\cite{ishihara-takahashi-2024-quantifying-memorization, takahashi-ishihara-2025-quantifying, zhang-etal-2024-pretraining}.
It is important to promote MIA research that is not limited to English.

\begin{table*}[t]
\centering
\setlength\tabcolsep{4pt}
\renewcommand{\arraystretch}{0.95}
\small
\begin{tabular}{llllll}
\toprule
Type & Method & AUC & Inference time (ratio) & FPR@95 & TPR@5 \\
\midrule
baseline & LOSS & 69.4 / 69.4 & 12s / 57s ($\times 4.75$) & 84.3 / 84.3 & 18.3 / 18.3 \\
& PPL/zlib & 69.8 / 69.8 & 12s / 57s ($\times 4.75$) & 80.2 / 80.2 & 14.5 / 14.5 \\\midrule
token distribution & Min-K\% Prob (K=0.1) & 67.2 / 67.2 & 12s / 57s ($\times 4.75$) & 83.5 / 83.3 & 17.3 / 17.3 \\
& Min-K\% Prob (K=0.2) & 69.3 / 69.3 & 12s / 57s ($\times 4.75$) & 82.3 / 82.3 & 22.0 / 22.0 \\
& Min-K\% Prob (K=0.3) & 70.1 / 70.1 & 12s / 57s ($\times 4.75$) & 82.3 / 82.3 & 19.6 / 19.6 \\
& Min-K\% Prob (K=0.5) & 69.7 / 69.7 & 12s / 57s ($\times 4.75$) & 82.5 / 82.5 & 18.1 / 18.1 \\
& Min-K\% Prob (K=0.8) & 69.5 / 69.5 & 12s / 57s ($\times 4.75$) & 84.3 / 84.3 & 18.1 / 18.3 \\
& Min-K\% Prob (K=1.0) & 69.4 / 69.4 & 12s / 57s ($\times 4.75$) & 84.3 / 84.3 & 18.3 / 18.3 \\
& DC-PDD & 67.4 / 67.4 & 12s / 57s ($\times 4.75$) & 84.8 / 84.8 & 12.4 / 12.4 \\\midrule
text alternation & Lowercase & 64.1 / 64.1 & 25s / 1m59s ($\times 4.76$) & 83.5 / 83.8 & 11.6 / 11.6 \\
& PAC & 73.3 / 73.4 & 1m17s / 6m24s ($\times 4.99$) & 82.3 / 77.9 & 27.6 / 24.3 \\
& ReCaLL & 90.7 / 90.3 & 55s / 2m10s ($\times 2.36$) & 28.5 / 34.7 & 50.4 / 48.8 \\
& Con-ReCall & 96.8 / 96.1 & 1m53s / 3m30s ($\times 1.86$) & 10.8 / 12.9 & 78.0 / 73.6 \\\midrule
black-box & SaMIA & 65.5 / 64.5 & 2h3m24s / 40h9m53s ($\times 19.5$) & 90.5 / 90.7 & 22.7 / 15.5 \\
\bottomrule
\end{tabular}
\caption{
    Comparison of Fast-MIA (left) and Transformers-based implementations (right) in terms of AUC, inference time, FPR@95, and TPR@5.
    Performance metrics such as AUC remain almost the same, while inference time is approximately five times faster.
}
\label{tab:comparison}
\end{table*}

\begin{table*}[t]
\centering
\setlength\tabcolsep{4pt}
\renewcommand{\arraystretch}{0.95}
\small
\begin{tabular}{llrrr}
\toprule
Type & Method & \multicolumn{2}{c}{Fast-MIA} & Transformers \\ \cmidrule{3-4}
& & w/ cache & w/o cache \\
\midrule
baseline & LOSS & 12s (1) & 12s (1) & 57s (1) \\
& PPL/zlib & 0s (0) & 12s (1) & 57s (1)  \\\midrule
token distribution & Min-K\% Prob & 0s (0) & 12s (1) & 57s (1) \\
& DC-PDD & 0s (0) & 12s (1) & 57s (1)\\\midrule
text alternation & Lowercase & 13s (1)  & 25s (2) 
& 1m59s (2) \\
& PAC & 1m5s (5) & 1m17s (6) & 6m24s (6) \\
& ReCaLL & 43s (1) & 55s (2) & 2m10s (2) \\
& Con-ReCall & 1m41s (2) & 1m53s (3) & 3m30s (3) \\\midrule
& Total & 3m54s (10) & 5m18s (17) & 17m51s (17) \\\bottomrule
\end{tabular}
\caption{
    Comparison of Fast-MIA with (w/) and without (w/o) cache, and Transformers-based implementations in terms of inference time (the number of inferences).
    We excluded SaMIA due to its extremely long inference time.
}
\label{tab:counts}
\end{table*}

\section{Evaluation}
\label{sec:evaluation}

The evaluation of Fast-MIA was conducted with two main objectives corresponding to key features:
\begin{itemize}
    \item To demonstrate that Fast-MIA achieves improvements in inference speed compared to naive Transformers implementations while ensuring correctness and reproducibility.
    \item To confirm that the total number of inferences is reduced compared to executing MIA methods individually.
\end{itemize}

\subsection{Experimental Setup}

To this end, we conducted comparative experiments using the same LLM model (LLaMA 30B~\cite{Touvron2023-qt}) and MIA settings on an NVIDIA A100 80GB GPU.
Fast-MIA was benchmarked against corresponding reference implementations using Transformers for all nine methods supported by Fast-MIA version 0.3.0.
The evaluation was performed on the WikiMIA dataset, with input token length set to 32.
We calculated the AUC, FPR@95, and TPR@5.
We did not include a comparison with existing toolkits such as those described in Table~\ref{tab:related}.
This is because available toolkits differ in supported methods, maintenance status, and backend versions, making direct end-to-end comparison difficult.

Each MIA method was executed by specifying only a single YAML configuration file, without additional scripting.
Methods that include hyperparameters require configuration.
$K$ in Min-K\% Prob was set to 0.1, 0.2, 0.3, 0.5, 0.8, and 1.0.
The original paper~\cite{shi2024detecting} reported that $K=0.2$ yielded the best results.
Note that the setting $K=1.0$ is consistent with LOSS.
In our settings, ReCaLL and Con-ReCall provided three examples as prompts, PAC was performed with five token swaps, and SaMIA used five outputs.
The complete configuration file used in Section~\ref{sec:evaluation} can be found in the Appendix~\ref{sec:configuration}.

\subsection{Evaluation Results}

As shown in Table~\ref{tab:comparison}, Fast-MIA reproduces the AUC scores from the Transformers-based implementation with negligible differences, while reducing the total computation time by approximately five times.
AUC is completely consistent in baseline methods and token distribution based methods.
In text alternation based and black-box methods, slight performance fluctuations occurred when generating text multiple times.
Particularly in SaMIA, substantial reductions in inference time were observed.
By replacing repeated generation loops with batched multi-output generation\footnote{In an intermediate Transformers-based implementation, replacing the per-sample generation loop with batched multi-output generation reduced the runtime to approximately 13 hours.} and leveraging vLLM-based inference, Fast-MIA reduced the runtime from 40h9m53s to 2h3m24s.

The cache mechanism reduced the inference time and the number of inferences in this experiment as shown in Table~\ref{tab:counts}.
Under cache-effective conditions, the intermediate results from the initial inference run with LOSS were reused.
As a result, inferences via token-distribution based methods became unnecessary.
For methods requiring multiple inferences, the number of inferences decreased by one each time.

Table~\ref{tab:counts} also serves as an ablation of the two efficiency mechanisms.
The comparison between the Transformers-based implementation and Fast-MIA without cache mainly reflects the effect of the vLLM-based batched inference backend, whereas the comparison between Fast-MIA with and without cache isolates the effect of cross-method caching.

\section{Conclusion}
\label{sec:conclusion}

We have introduced Fast-MIA, a Python library that enables efficient and scalable evaluation of MIA against LLMs.
Against the background of high computational costs and the lack of standardized implementations, Fast-MIA provides high-throughput batch inference on vLLM and cross-method caching architecture, enabling rapid comparative evaluation across diverse methods.
The experiment demonstrated that Fast-MIA can reduce inference time with small performance changes.
We release Fast-MIA as an open-source tool to accelerate research on privacy, memorization, and evaluation of LLMs.
We anticipate that this library serves as a foundation for future methodological and empirical advancements.

\section*{Limitations}

While Fast-MIA offers a scalable and extensible evaluation platform, the current implementation has several limitations:
\begin{itemize}
  \item \textbf{Number of implemented methods}: Numerous methods have been proposed for membership inference on LLMs. While Fast-MIA covers representative approaches, there remains room for further implementation. Methods targeting entire datasets have also not yet been implemented.
  \item \textbf{Model diversity}: Fast-MIA currently supports models compatible with vLLM, with open-weight autoregressive language models benefiting most significantly. Models based on different architectures (e.g., encoder-only or encoder-decoder) are not yet sufficiently supported. Although black-box methods could potentially be applied to closed APIs, they are not supported in the current implementation.
  \item \textbf{Broader customization:} Although Fast-MIA supports custom MIA methods through a modular interface, other extensions such as user-defined evaluation metrics, customized report formats, and new output schemas are not yet fully configurable through YAML configuration files. Supporting these broader customization needs without modifying the internal evaluation or reporting pipeline is an important direction for future work.
\end{itemize}
Addressing these limitations is part of our ongoing work.
We welcome community contributions to broaden the scope of attack types, model families, and other areas supported by Fast-MIA.

Furthermore, since the primary purpose of this paper is to demonstrate Fast-MIA, the experiments are not exhaustive.
Our experiments focus on one representative model, dataset, text length, and hardware setting to validate the correctness and efficiency of the library under a controlled setup.
Although the acceleration mechanisms are model-agnostic for vLLM-compatible autoregressive models, the exact speedup may vary depending on model size, sequence length, hardware, and decoding configuration.
Broader evaluations across model families, dataset domains, context lengths, and hardware environments are left for future work.

\section*{Ethics/Broader Impact}

Fast-MIA is designed to advance the understanding of privacy risks in LLMs by enabling reproducible and standardized evaluation of MIAs.
While the methods implemented in the library can potentially be used to identify whether a given text was part of a model’s training data, the primary purpose of this work is to support responsible research into memorization and data leakage.

We emphasize that the library does not include pre-trained MIA models, or automatic data scraping routines.
Instead, it focuses on benchmarking known MIA strategies under controlled experimental settings.
We believe that enabling reproducible and transparent evaluation of MIA risk is a necessary step toward the development of safer and more privacy-preserving LLMs.

Fast-MIA is released under an open-source license, but we encourage users to follow ethical research practices, including proper citation and responsible data handling.
We position Fast-MIA as a platform to promote robust and responsible privacy-aware development of LLMs, rather than as a tool for exploitation.

\appendix

\section{Configuration File Used in Section 4}
\label{sec:configuration}

The complete YAML configuration file used in Section~\ref{sec:evaluation} is as follows.

\begin{quote}
\footnotesize
\begin{verbatim}
# Model settings
model:
  model_id: "huggyllama/llama-30b"
  trust_remote_code: true
  max_num_seqs: 256

# sampling_parameters settings
sampling_parameters:
  max_tokens: 1
  prompt_logprobs: 0
  temperature: 0.0
  top_p: 1.0

# Data settings
data:
  # Basic settings
  data_path: "swj0419/WikiMIA"
  # Path to data file or huggingface dataset name
  format: "huggingface"
  # Data format
  # (csv, jsonl, json, parquet, huggingface)

  # Column name settings
  text_column: "input"
  # Name of text column
  label_column: "label"
  # Name of label column

  # text split settings
  text_length: 32
  # Number of words to split
  # (for WikiMIA dataset: 32, 64, 128, or 256)

  # Language
  space_delimited_language: true

# Evaluation methods
methods:
  - type: "loss"
    params: {}
  - type: "lower"
    params: {}
  - type: "zlib"
    params: {}
  - type: "mink"
    params:
      ratio: 0.1
  - type: "mink"
    params:
      ratio: 0.2
  - type: "mink"
    params:
      ratio: 0.3
  - type: "mink"
    params:
      ratio: 0.5
  - type: "mink"
    params:
      ratio: 0.8
  - type: "mink"
    params:
      ratio: 1.0
  - type: "recall"
    params:
      num_shots: 3
      pass_window: false
  - type: "conrecall"
    params:
      num_shots: 3
      pass_window: false
      gamma: 0.5
  - type: "pac"
    params:
      alpha: 0.3
      N: 5
  - type: "samia"
    params:
      num_samples: 5
      prefix_ratio: 0.5
      zlib: true
  - type: "dcpdd"
    params:
      file_num: 15
      max_token_length: 1024
      alpha: 0.01

# Output settings
output_dir: "./results" 
\end{verbatim}
\end{quote}


\begin{thebibliography}{31}
\providecommand{\natexlab}[1]{#1}

\bibitem[{Carlini et~al.(2022)Carlini, Chien, Nasr, Song, Terzis, and Tramèr}]{Carlini2022-rx}
Nicholas Carlini, Steve Chien, Milad Nasr, Shuang Song, Andreas Terzis, and Florian Tramèr. 2022.
\newblock \href {https://doi.ieeecomputersociety.org/10.1109/SP46214.2022.9833649} {Membership inference attacks from first principles}.
\newblock In \emph{2022 IEEE Symposium on Security and Privacy (SP)}, pages 1897--1914. IEEE.

\bibitem[{Carlini et~al.(2021)Carlini, Tram{\`e}r, Wallace, Jagielski, Herbert-Voss, Lee, Roberts, Brown, Song, Erlingsson, Oprea, and Raffel}]{274574}
Nicholas Carlini, Florian Tram{\`e}r, Eric Wallace, Matthew Jagielski, Ariel Herbert-Voss, Katherine Lee, Adam Roberts, Tom Brown, Dawn Song, {\'U}lfar Erlingsson, Alina Oprea, and Colin Raffel. 2021.
\newblock \href {https://www.usenix.org/conference/usenixsecurity21/presentation/carlini-extracting} {Extracting training data from large language models}.
\newblock In \emph{30th USENIX Security Symposium (USENIX Security 21)}, pages 2633--2650. USENIX Association.

\bibitem[{Chen et~al.(2025)Chen, Han, and Miyao}]{chen-etal-2025-statistical}
Bowen Chen, Namgi Han, and Yusuke Miyao. 2025.
\newblock \href {https://doi.org/10.18653/v1/2025.acl-long.1114} {A statistical and multi-perspective revisiting of the membership inference attack in large language models}.
\newblock In \emph{Proceedings of the 63rd Annual Meeting of the Association for Computational Linguistics (Volume 1: Long Papers)}, pages 22854--22874, Vienna, Austria. Association for Computational Linguistics.

\bibitem[{Chen et~al.(2026)Chen, Du, Zhang, Kundu, Fleming, Ribeiro, and Li}]{Chen2026-om}
Yuetian Chen, Yuntao Du, Kaiyuan Zhang, Ashish Kundu, Charles Fleming, Bruno Ribeiro, and Ninghui Li. 2026.
\newblock \href {https://arxiv.org/abs/2601.02751} {Window-based membership inference attacks against fine-tuned large language models}.
\newblock \emph{arXiv [cs.CL]}.

\bibitem[{Dong et~al.(2024)Dong, Jiang, Liu, Jin, Gu, Yang, and Li}]{dong-etal-2024-generalization}
Yihong Dong, Xue Jiang, Huanyu Liu, Zhi Jin, Bin Gu, Mengfei Yang, and Ge~Li. 2024.
\newblock \href {https://doi.org/10.18653/v1/2024.findings-acl.716} {Generalization or memorization: Data contamination and trustworthy evaluation for large language models}.
\newblock In \emph{Findings of the Association for Computational Linguistics: ACL 2024}, pages 12039--12050, Bangkok, Thailand. Association for Computational Linguistics.

\bibitem[{Duan et~al.(2024)Duan, Suri, Mireshghallah, Min, Shi, Zettlemoyer, Tsvetkov, Choi, Evans, and Hajishirzi}]{duan2024do}
Michael Duan, Anshuman Suri, Niloofar Mireshghallah, Sewon Min, Weijia Shi, Luke Zettlemoyer, Yulia Tsvetkov, Yejin Choi, David Evans, and Hannaneh Hajishirzi. 2024.
\newblock \href {https://openreview.net/forum?id=av0D19pSkU} {Do membership inference attacks work on large language models?}
\newblock In \emph{First Conference on Language Modeling}.

\bibitem[{Hu et~al.(2022)Hu, Shen, Wallis, Allen-Zhu, Li, Wang, Wang, and Chen}]{hu2022lora}
Edward~J Hu, Yelong Shen, Phillip Wallis, Zeyuan Allen-Zhu, Yuanzhi Li, Shean Wang, Lu~Wang, and Weizhu Chen. 2022.
\newblock \href {https://openreview.net/forum?id=nZeVKeeFYf9} {Lo{RA}: Low-rank adaptation of large language models}.
\newblock In \emph{International Conference on Learning Representations}.

\bibitem[{Ishihara(2023)}]{ishihara-2023-training}
Shotaro Ishihara. 2023.
\newblock \href {https://doi.org/10.18653/v1/2023.trustnlp-1.23} {Training data extraction from pre-trained language models: A survey}.
\newblock In \emph{Proceedings of the 3rd Workshop on Trustworthy Natural Language Processing (TrustNLP 2023)}, pages 260--275, Toronto, Canada. Association for Computational Linguistics.

\bibitem[{Ishihara and Takahashi(2024)}]{ishihara-takahashi-2024-quantifying-memorization}
Shotaro Ishihara and Hiromu Takahashi. 2024.
\newblock \href {https://doi.org/10.18653/v1/2024.inlg-main.14} {Quantifying memorization and detecting training data of pre-trained language models using {J}apanese newspaper}.
\newblock In \emph{Proceedings of the 17th International Natural Language Generation Conference}, pages 165--179, Tokyo, Japan. Association for Computational Linguistics.

\bibitem[{Kaneko et~al.(2025)Kaneko, Ma, Wata, and Okazaki}]{kaneko-etal-2025-sampling}
Masahiro Kaneko, Youmi Ma, Yuki Wata, and Naoaki Okazaki. 2025.
\newblock \href {https://doi.org/10.18653/v1/2025.findings-acl.465} {Sampling-based pseudo-likelihood for membership inference attacks}.
\newblock In \emph{Findings of the Association for Computational Linguistics: ACL 2025}, pages 8894--8907, Vienna, Austria. Association for Computational Linguistics.

\bibitem[{Koike et~al.(2025)Koike, Dugan, Kaneko, Callison-Burch, and Okazaki}]{Koike2025-hv}
Ryuto Koike, Liam Dugan, Masahiro Kaneko, Chris Callison-Burch, and Naoaki Okazaki. 2025.
\newblock \href {https://arxiv.org/abs/2510.19492} {Machine text detectors are membership inference attacks}.
\newblock \emph{arXiv [cs.CL]}.

\bibitem[{Kwon et~al.(2023)Kwon, Li, Zhuang, Sheng, Zheng, Yu, Gonzalez, Zhang, and Stoica}]{10.1145/3600006.3613165}
Woosuk Kwon, Zhuohan Li, Siyuan Zhuang, Ying Sheng, Lianmin Zheng, Cody~Hao Yu, Joseph Gonzalez, Hao Zhang, and Ion Stoica. 2023.
\newblock \href {https://doi.org/10.1145/3600006.3613165} {Efficient memory management for large language model serving with pagedattention}.
\newblock In \emph{Proceedings of the 29th Symposium on Operating Systems Principles}, SOSP '23, page 611–626, New York, NY, USA. Association for Computing Machinery.

\bibitem[{Lee et~al.(2023)Lee, Le, Chen, and Lee}]{10.1145/3543507.3583199}
Jooyoung Lee, Thai Le, Jinghui Chen, and Dongwon Lee. 2023.
\newblock \href {https://doi.org/10.1145/3543507.3583199} {Do language models plagiarize?}
\newblock In \emph{Proceedings of the ACM Web Conference 2023}, WWW '23, page 3637–3647, New York, NY, USA. Association for Computing Machinery.

\bibitem[{Magar and Schwartz(2022)}]{magar-schwartz-2022-data}
Inbal Magar and Roy Schwartz. 2022.
\newblock \href {https://doi.org/10.18653/v1/2022.acl-short.18} {Data contamination: From memorization to exploitation}.
\newblock In \emph{Proceedings of the 60th Annual Meeting of the Association for Computational Linguistics (Volume 2: Short Papers)}, pages 157--165, Dublin, Ireland. Association for Computational Linguistics.

\bibitem[{Maini et~al.(2024)Maini, Jia, Papernot, and Dziedzic}]{NEURIPS2024_e01519b4}
Pratyush Maini, Hengrui Jia, Nicolas Papernot, and Adam Dziedzic. 2024.
\newblock \href {https://doi.org/10.52202/079017-3941} {{LLM} dataset inference: Did you train on my dataset?}
\newblock In \emph{Advances in Neural Information Processing Systems}, volume~37, pages 124069--124092. Curran Associates, Inc.

\bibitem[{Meeus et~al.(2025)Meeus, Shilov, Jain, Faysse, Rei, and de~Montjoye}]{10992321}
Matthieu Meeus, Igor Shilov, Shubham Jain, Manuel Faysse, Marek Rei, and Yves-Alexandre de~Montjoye. 2025.
\newblock \href {https://doi.org/10.1109/SaTML64287.2025.00028} {{ SoK: Membership Inference Attacks on LLMs are Rushing Nowhere (and How to Fix It) }}.
\newblock In \emph{2025 IEEE Conference on Secure and Trustworthy Machine Learning (SaTML)}, pages 385--401, Los Alamitos, CA, USA. IEEE Computer Society.

\bibitem[{Murakonda and Shokri(2020)}]{Murakonda2020-yr}
Sasi~Kumar Murakonda and Reza Shokri. 2020.
\newblock \href {https://arxiv.org/abs/2007.09339} {{ML} privacy meter: Aiding regulatory compliance by quantifying the privacy risks of machine learning}.
\newblock \emph{arXiv [cs.CR]}.

\bibitem[{Puerto et~al.(2025)Puerto, Gubri, Yun, and Oh}]{puerto-etal-2025-scaling}
Haritz Puerto, Martin Gubri, Sangdoo Yun, and Seong~Joon Oh. 2025.
\newblock \href {https://doi.org/10.18653/v1/2025.findings-naacl.234} {Scaling up membership inference: When and how attacks succeed on large language models}.
\newblock In \emph{Findings of the Association for Computational Linguistics: NAACL 2025}, pages 4165--4182, Albuquerque, New Mexico. Association for Computational Linguistics.

\bibitem[{Radford et~al.(2019)Radford, Wu, Child, Luan, Amodei, and Sutskever}]{radford2019language}
Alec Radford, Jeff Wu, Rewon Child, David Luan, Dario Amodei, and Ilya Sutskever. 2019.
\newblock \href {https://cdn.openai.com/better-language-models/language_models_are_unsupervised_multitask_learners.pdf} {Language models are unsupervised multitask learners}.

\bibitem[{Ravaut et~al.(2025)Ravaut, Ding, Jiao, Chen, Li, Zhao, Qin, Xiong, and Joty}]{Ravaut2024-eh}
Mathieu Ravaut, Bosheng Ding, Fangkai Jiao, Hailin Chen, Xingxuan Li, Ruochen Zhao, Chengwei Qin, Caiming Xiong, and Shafiq Joty. 2025.
\newblock \href {https://openreview.net/forum?id=SxNMjbtdFm} {A comprehensive survey of contamination detection methods in large language models}.
\newblock \emph{Transactions on Machine Learning Research}.

\bibitem[{Shi et~al.(2024)Shi, Ajith, Xia, Huang, Liu, Blevins, Chen, and Zettlemoyer}]{shi2024detecting}
Weijia Shi, Anirudh Ajith, Mengzhou Xia, Yangsibo Huang, Daogao Liu, Terra Blevins, Danqi Chen, and Luke Zettlemoyer. 2024.
\newblock \href {https://openreview.net/forum?id=zWqr3MQuNs} {Detecting pretraining data from large language models}.
\newblock In \emph{The Twelfth International Conference on Learning Representations}.

\bibitem[{Shokri et~al.(2017)Shokri, Stronati, Song, and Shmatikov}]{7958568}
Reza Shokri, Marco Stronati, Congzheng Song, and Vitaly Shmatikov. 2017.
\newblock \href {https://doi.org/10.1109/SP.2017.41} {Membership inference attacks against machine learning models}.
\newblock In \emph{2017 IEEE Symposium on Security and Privacy (SP)}, pages 3--18.

\bibitem[{Takahashi and Ishihara(2025)}]{takahashi-ishihara-2025-quantifying}
Hiromu Takahashi and Shotaro Ishihara. 2025.
\newblock \href {https://doi.org/10.18653/v1/2025.l2m2-1.8} {Quantifying memorization in continual pre-training with {J}apanese general or industry-specific corpora}.
\newblock In \emph{Proceedings of the First Workshop on Large Language Model Memorization (L2M2)}, pages 95--105, Vienna, Austria. Association for Computational Linguistics.

\bibitem[{Touvron et~al.(2023)Touvron, Lavril, Izacard, Martinet, Lachaux, Lacroix, Rozière, Goyal, Hambro, Azhar, Rodriguez, Joulin, Grave, and Lample}]{Touvron2023-qt}
Hugo Touvron, Thibaut Lavril, Gautier Izacard, Xavier Martinet, Marie-Anne Lachaux, Timothée Lacroix, Baptiste Rozière, Naman Goyal, Eric Hambro, Faisal Azhar, Aurelien Rodriguez, Armand Joulin, Edouard Grave, and Guillaume Lample. 2023.
\newblock \href {https://doi.org/10.48550/arXiv.2302.13971} {{LLaMA}: Open and efficient foundation language models}.
\newblock \emph{arXiv [cs.CL]}.

\bibitem[{Wang et~al.(2025)Wang, Wang, Hooi, Cai, Peng, and Chang}]{wang-etal-2025-con}
Cheng Wang, Yiwei Wang, Bryan Hooi, Yujun Cai, Nanyun Peng, and Kai-Wei Chang. 2025.
\newblock \href {https://aclanthology.org/2025.coling-main.68/} {Con-{R}e{C}all: Detecting pre-training data in {LLM}s via contrastive decoding}.
\newblock In \emph{Proceedings of the 31st International Conference on Computational Linguistics}, pages 1013--1026, Abu Dhabi, UAE. Association for Computational Linguistics.

\bibitem[{Wolf et~al.(2020)Wolf, Debut, Sanh, Chaumond, Delangue, Moi, Cistac, Rault, Louf, Funtowicz, Davison, Shleifer, von Platen, Ma, Jernite, Plu, Xu, Le~Scao, Gugger, Drame, Lhoest, and Rush}]{wolf-etal-2020-transformers}
Thomas Wolf, Lysandre Debut, Victor Sanh, Julien Chaumond, Clement Delangue, Anthony Moi, Pierric Cistac, Tim Rault, Remi Louf, Morgan Funtowicz, Joe Davison, Sam Shleifer, Patrick von Platen, Clara Ma, Yacine Jernite, Julien Plu, Canwen Xu, Teven Le~Scao, Sylvain Gugger, and 3 others. 2020.
\newblock \href {https://doi.org/10.18653/v1/2020.emnlp-demos.6} {Transformers: State-of-the-art natural language processing}.
\newblock In \emph{Proceedings of the 2020 Conference on Empirical Methods in Natural Language Processing: System Demonstrations}, pages 38--45, Online. Association for Computational Linguistics.

\bibitem[{Wu and Cao(2025)}]{Wu2025-ar}
Hengyu Wu and Yang Cao. 2025.
\newblock \href {https://arxiv.org/abs/2503.19338} {Membership inference attacks on large-scale models: A survey}.
\newblock \emph{arXiv [cs.LG]}.

\bibitem[{Xie et~al.(2024)Xie, Wang, Huang, Zhang, Ge, Pei, Gong, and Dhingra}]{xie-etal-2024-recall}
Roy Xie, Junlin Wang, Ruomin Huang, Minxing Zhang, Rong Ge, Jian Pei, Neil~Zhenqiang Gong, and Bhuwan Dhingra. 2024.
\newblock \href {https://doi.org/10.18653/v1/2024.emnlp-main.493} {{R}e{C}a{LL}: Membership inference via relative conditional log-likelihoods}.
\newblock In \emph{Proceedings of the 2024 Conference on Empirical Methods in Natural Language Processing}, pages 8671--8689, Miami, Florida, USA. Association for Computational Linguistics.

\bibitem[{Ye et~al.(2024)Ye, Hu, Li, Wang, Chen, and Zhao}]{ye-etal-2024-data}
Wentao Ye, Jiaqi Hu, Liyao Li, Haobo Wang, Gang Chen, and Junbo Zhao. 2024.
\newblock \href {https://doi.org/10.18653/v1/2024.findings-acl.644} {Data contamination calibration for black-box {LLM}s}.
\newblock In \emph{Findings of the Association for Computational Linguistics: ACL 2024}, pages 10845--10861, Bangkok, Thailand. Association for Computational Linguistics.

\bibitem[{Yeom et~al.(2018)Yeom, Giacomelli, Fredrikson, and Jha}]{8429311}
Samuel Yeom, Irene Giacomelli, Matt Fredrikson, and Somesh Jha. 2018.
\newblock \href {https://doi.org/10.1109/CSF.2018.00027} {Privacy risk in machine learning: Analyzing the connection to overfitting}.
\newblock In \emph{2018 IEEE 31st Computer Security Foundations Symposium (CSF)}, pages 268--282.

\bibitem[{Zhang et~al.(2024)Zhang, Zhang, Guo, de~Rijke, Fan, and Cheng}]{zhang-etal-2024-pretraining}
Weichao Zhang, Ruqing Zhang, Jiafeng Guo, Maarten de~Rijke, Yixing Fan, and Xueqi Cheng. 2024.
\newblock \href {https://doi.org/10.18653/v1/2024.emnlp-main.300} {Pretraining data detection for large language models: A divergence-based calibration method}.
\newblock In \emph{Proceedings of the 2024 Conference on Empirical Methods in Natural Language Processing}, pages 5263--5274, Miami, Florida, USA. Association for Computational Linguistics.

\end{thebibliography}
\end{document}